\documentclass[prc,preprint]{revtex4}
\usepackage{epsfig}
\begin{document}
\flushbottom

\title{Calculation of heavy ion $e^+ e^-$ pair production to all orders in
$Z \alpha$}
\author{A. J. Baltz}
\affiliation{Physics Department,
Brookhaven National Laboratory,
Upton, New York 11973}
\date{April 21, 2005}
%\date{\today}

\begin{abstract}
The heavy ion cross section for continuum $e^+ e^-$ pair production has
been calculated to all orders in $Z \alpha$.  The formula resulting from an
exact solution of the semiclassical Dirac equation in the ultrarelativistic
limit is evaluated numerically.  An energy dependent spatial cutoff of
the heavy ion potential is utilized, leading to an
exact formula agreeing with the known perturbative formula in the
ultrarelativistic, perturbative limit.  Cross sections and sample momentum
distributions are evaluated for heavy ion beams at SPS, RHIC, and LHC energies.
$e^+ e^-$ pair production
probabilities are found to be reduced from perturbation theory with
increasing charge of the colliding heavy ions and for all energy and
momentum regions investigated.
\end{abstract}
\maketitle
%\nopagebreak
%\twocolumn
%\narrowtext
\section{Introduction}
The calculation of heavy ion induced continuum $e^+ e^-$ pair production to
all orders in $Z \alpha$ is of continuing interest because up to now it has
only been carried out for total cross sections and in a limiting
approximation\cite{serb,lm1,lm2}.  Recent progress on this
topic began with the realization that in an appropriate gauge\cite{brbw},
the electromagnetic field of a relativistic heavy ion is to a very good
approximation a delta function in the direction of motion of the heavy ion
times the two dimensional solution of Maxwell's equations in the transverse
direction\cite{ajb1}.  This realization led to an exact solution of the
appropriate Dirac equation for excitation of bound-electron positron
pairs and a predicted reduction from perturbation theory of a little less
than 10\% for Au + Au at RHIC \cite{ajb2}.  This reduction can be identified
as a Coulomb correction to bound-electron pair production.

It soon followed that an analytical solution of the Dirac equation was obtained
independently and practically simultaneously by two
different collaborations\cite{sw1,bmc,sw2} for the analagous case of continuum
$e^+ e^-$ pair production induced by the corresponding countermoving delta
function potentials produced by ultrarelativistic heavy ions in a collider
such as RHIC.  An extended discussion and reanalyis of this solution, with
comments on early parallel work in the literature, shortly followed\cite{eg}.
Baltz and McLerran\cite{bmc} noted the apparent agreement of
the obtained amplitude with that of perturbation theory even for large $Z$.
Segev and Wells\cite{sw2} further noted the perturbative scaling with
$Z_1^2 Z_2^2$ seen in CERN SPS data\cite{vd}. These data were obtained from
reactions of 160 GeV/nucleon Pb ions on C, Al, Pa, and Au targets
as well as from 200 Gev/nucleon S ions on the same C, Al, Pa, and Au targets.
The group presenting the CERN data, Vane et al., stated their findings
in summary: ``Cross sections
scale as the product of the squares of the projectile
and target nuclear charges.''  On the other hand, it is well known that
photoproduction of $e^+ e^-$ pairs on a heavy target shows a negative
(Coulomb) correction proportional to $Z^2$ that is well described by the
Bethe-Maximon theory\cite{bem}.

Several authors subsequently argued\cite{serb,lm1,lm2} that a correct
regularization of the exact Dirac equation amplitude should lead to
a reduction of the total cross section for pair production from
perturbation theory, the so called Coulomb corrections.  The first
analysis was done in a Weizsacker-Williams approximation\cite{serb}.
Subsequently, Lee and Milstein computed\cite{lm1,lm2} the total cross
section for $e^+ e^-$ pairs using approximations to the exact amplitude
that led to a higher order correction to the well known Landau-Lifshitz
expression\cite{ll}.  In a previous paper\cite{ajb3} I have tried to
explicate the Lee and Milstein approximate results and argued their
qualitative correctness.

In the present paper I undertake the full numerical calculation of
electromagnetically induced ultrarelativistic heavy ion electron-positron
pair production.  I utilize a cross section formula derived from
the exact solution of the Dirac equation with an appropriate energy
dependent cutoff of the transversely eikonalized potential employed.  

\section{Cross Sections with Higher order Coulomb corrections}

For production of continuum pairs in an ultrarelativistic heavy ion reaction
one may work in a frame of two countermoving heavy ions with the same
relativistic $\gamma$, and the electromagnetic interaction arising from them
goes to the limit of two $\delta$ function potentials
\begin{equation}
V(\mbox{\boldmath $ \rho$},z,t)
=\delta(z - t) (1-\alpha_z) \Lambda^-(\mbox{\boldmath $ \rho$})
+\delta(z + t) (1+\alpha_z) \Lambda^+(\mbox{\boldmath $ \rho$}) 
\end{equation}
where
\begin{equation}
\Lambda^{\pm}(\mbox{\boldmath $ \rho$}) = - Z \alpha 
\ln {(\mbox{\boldmath $ \rho$} \pm {\bf b}/2)^2 \over (b/2)^2}.
\end{equation}

The previously derived semiclassical amplitude for
electron-positron pair production\cite{sw1,bmc,sw2,eg} written in the
notation of Lee and Milstein\cite{lm1} takes the form
\begin{equation}
M(p,q) = \int { d^2 k \over (2 \pi)^2 } \exp [i\, {\bf k} \cdot {\bf b}]
{\cal M}({\bf k}) F_B({\bf k})
F_A({\bf q_{\perp} + p_{\perp} - k})
\end{equation}
where $p$ and $q$ are the four-momenta of the produced electron and positron
respectively, $p_{\pm}=p_0 \pm p_z$, $q_{\pm}=q_0 \pm p_z$,
$\gamma_{\pm} = \gamma_0 \pm \gamma_z$,
$\mbox{\boldmath $\alpha$}=\gamma_0 \mbox{\boldmath $\gamma$}$,
${\bf k}$ is an intermediate transverse momentum transfer from the ion to be
integrated over,
\begin{eqnarray}
{\cal M}({\bf k}) &=& \bar{u}(p) {\mbox{\boldmath $\alpha$} \cdot
({\bf k -p_{\perp}})
+ \gamma_0 m  \over  -p_+ q_- -({\bf k - p_{\perp}})^2 - m^2 + i \epsilon}
\gamma_- u(-q)
 \nonumber \\
&\ & +  \bar{u}(p) {- \mbox{\boldmath $\alpha$} \cdot ({\bf k -q_{\perp}})
+ \gamma_0 m  \over  -p_-q_+ -({\bf k - q_{\perp}})^2 - m^2 + i \epsilon}
\gamma_+ u(-q),
\end{eqnarray}
and the effect of the potential Eq. (1-2) is contained in integrals, 
$F_B$ and $F_A$, over the transverse spatial coordinates\cite{sw1,bmc,sw2,eg},
\begin{equation}
F({\bf k}) = \int d^2 \rho\, 
\exp [-i\, {\bf k} \cdot \mbox{\boldmath $ \rho$}] 
\{ \exp [-i  2 Z \alpha \ln {\rho}] - 1 \} .
\end{equation}
$F({\bf k})$ has to be regularized or cut off at large $\rho$.
How it is regularized
is the key to understanding Coulomb corrections.  

Although, as has been pointed out\cite{bg}, the derived exact semiclassical
Dirac amplitude is not simply the exact amplitude for the excitation of a
particular (correlated) electron-positron pair, there are observables,
such as the total pair production cross section, that can be
constructed straightforwardly from this derived
amplitude\cite{read,rmgs,mom,obe}.  This point
has a long history of discussion in the
literature\cite{Baur90,Best92,HenckenTB95a,Aste02}.
The exact amplitude for a correlated electron-positron pair will not be
treated here.  The point is that exact solution of the semi-classical Dirac
equation may be used to compute the inclusive
average number of pairs --- not an exclusive amplitude for a particular pair.
Calculating the exact exclusive amplitude to all orders in $Z \alpha$
is not easily tractable due to need for Feynman propagators\cite{bg}.
The possibility of solutions of the semi-classical Dirac equation is
connected to the retarded propagators involved.  In this paper we do not
consider the exclusive (Feynman propagator) amplitude at all.  We concentrate
on observables that {\it can} be constructed from the above
amplitude and investigate the Coulomb corrections contained in them.

In a previous article\cite{ajb3} I have discussed these matters in more detail.
There
the uncorrelated cross section expressions for $d \sigma(p)$, $d \sigma(q)$,
and $\sigma_T$ were presented,
\begin{equation}
d \sigma(p) = \int {m\, d^3 q  \over (2 \pi)^3 \epsilon_q }
\int { d^2 k \over (2 \pi)^2 } \vert {\cal M}({\bf k}) \vert^2
\vert F_A({\bf q_{\perp} + p_{\perp} - k}) \vert^2 \vert F_B({\bf k}) \vert^2, 
\end{equation}
\begin{equation}
d \sigma(q) = \int {m\, d^3 p  \over (2 \pi)^3 \epsilon_p }
\int { d^2 k \over (2 \pi)^2 } \vert {\cal M}({\bf k}) \vert^2
\vert F_A({\bf q_{\perp} + p_{\perp} - k}) \vert^2 \vert F_B({\bf k}) \vert^2, 
\end{equation}
\begin{equation}
\sigma_T = \int {m^2 d^3 p\, d^3 q  \over (2 \pi)^6
\epsilon_p \epsilon_q }
\int { d^2 k \over (2 \pi)^2 } \vert {\cal M}({\bf k}) \vert^2
\vert F_A({\bf q_{\perp} + p_{\perp} - k}) \vert^2 \vert F_B({\bf k}) \vert^2.
\end{equation}
$d \sigma(p)$ is the cross section for an electron of momentum $(p)$ where
the state of the positron is unspecified.  Likewise $d \sigma(q)$ is the
cross section for a positron of momentum $(q)$ with
the state of the electron unspecified.
Note that $\sigma_T$ corresponds to a peculiar type of inclusive cross section
which we should call the ``number weighted total cross section'',
\begin{equation}
\sigma_T = \int d^2  b N = \int d^2 b \sum_{n=1}^{\infty} n P_n(b),
\end{equation}
in contrast to the usual definition of an inclusive total cross section
$\sigma_I$ for pair production,
\begin{equation}
\sigma_I = \int d^2 b \sum_{n=1}^{\infty} P_n(b).
\end{equation}

In Eqs. (6-8), it is assumed that the sums have been taken
over the
electron and positron polarizations in $\vert {\cal M}({\bf k}) \vert^2$.
Taking traces with the aid of the computer program FORM\cite{form} one obtains
\begin{eqnarray}
&&\vert{\cal M}({\bf k})\vert^2\  =\ \ {2 p_+ q_-[({\bf k -p_{\perp}})^2 + m^2]
\over m^2 [p_+ q_- +({\bf k - p_{\perp}})^2 + m^2]^2}
\ \ + \ \ {2 p_- q_+ [ ({\bf k -q_{\perp}})^2 + m^2]
\over m^2 [p_- q_+ +({\bf k - q_{\perp}})^2 + m^2]^2}
 \nonumber \\
&&+
{ 4[{\bf k \cdot p_{\perp}} q_+ q_- + {\bf k \cdot q_{\perp}} p_+ p_-
-2 {\bf k \cdot p_{\perp} k \cdot q_{\perp}}
+ k^2 ({\bf p_{\perp} \cdot q_{\perp}} - m^2) - p_+ p_- q_+ q_- ]   
\over m^2 [p_+ q_- +({\bf k - p_{\perp}})^2 + m^2]
[p_- q_+ +({\bf k - q_{\perp}})^2 + m^2]}.
\end{eqnarray}
This expression exhibits the expected property that
$\vert{\cal M}({\bf k})\vert^2$ vanishes as $\bf k$ goes to zero;
the positive squares of the direct and crossed amplitudes
(terms one and two) are cancelled by the negative product of direct and crossed
amplitudes of term three.  These background terms can be subtracted off
analytically from the expression for $\vert{\cal M}({\bf k})\vert^2$ to
obtain an expression exhibiting only terms dependent on $\bf k$ in the
numerators:
\begin{eqnarray}
\vert{\cal M}({\bf k})\vert^2 &=&{2 D_1^2 \eta_{11} - 2 A_{11}
(2 D_1 \beta_1 + \beta_1^2) \over m^2 D_1^2 (D_1 + \beta_1)^2}
+ {2 D_2^2 \eta_{22} - 2 A_{22}
(2 D_2 \beta_2 + \beta_2^2) \over m^2 D_2^2 (D_2 + \beta_2)^2}
\nonumber \\
&+& 4 {D_1 D_2 \eta_{12} - A_{12} ( D_2 \beta_1
+ \ D_1 \beta_2 + \beta_1 \beta_2 ) \over m^2 D_1 D_2 (D_1 + \beta_1) 
(D_2 + \beta_2)}
\end{eqnarray}
where
\begin{eqnarray}
A_{11}&=&p_+ q_-  ( {\bf p_{\perp}}^2 + m^2 )\ \ \ \ \ \ \ \ \ \ \ \ \ \ \ 
A_{22}=p_- q_+  ( {\bf q_{\perp}}^2 + m^2 ) 
\nonumber \\
A_{12}&=&-({\bf p_{\perp}}^2 + m^2 )({\bf q_{\perp}}^2 + m^2) 
\nonumber \\
D_1&=&p_+ q_- + {\bf p_{\perp}}^2 + m^2\ \ \ \ \ \ \ \ \ \ \ \ \ \ 
D_2=p_- q_+ + {\bf q_{\perp}}^2 + m^2
\nonumber \\
\beta_1&=&-2{\bf k \cdot p_{\perp}} + k^2\ \ \ \ \ \ \ \ \ \ \ \ \ \ 
\beta_2=-2{\bf k \cdot q_{\perp}} + k^2
\nonumber \\
\eta_{11}&=& p_+ q_-  \beta_1\ \ \ \ \ \ \ \ \ \ \ \ \ \ \ \ \ \ \ \ \ \ 
\eta_{22}= p_- q_+ \beta_2
\nonumber \\
\eta_{12}&=&{\bf k \cdot p_{\perp}} q_+ q_- + {\bf k \cdot q_{\perp}} p_+ p_-
-2 {\bf k \cdot p_{\perp} k \cdot q_{\perp}}
+ k^2 ({\bf p_{\perp} \cdot q_{\perp}} - m^2).
\end{eqnarray}
Every term in the numerators now has at least a linear dependence on $k$.
This subtraction turned out to be necessary to limit roundoff error in
calculations at the highest beam energies such as LHC.

If one merely regularizes
the integral itself at large $\rho$ one obtains\cite{bmc,sw2,eg} 
apart from a trivial phase
\begin{equation}
F({\bf k}) = {4 \pi \alpha Z \over k^{2 - 2 i \alpha Z} }.
\end{equation}
Then all the higher order $Z \alpha$ effects in $M(p,q)$ are contained only
in the phase of the denominator of Eq. (14).  Then, since the cross sections
Eq. (6-8) go as $\vert F({\bf k}) \vert^2 $ the phase falls out of the
problem and it directly
follows that calculable observables are equal to perturbative results.
However, in this approach a lower $k$ cutoff at some $\omega / \gamma$ has
to be put in by hand to obtain dependence on the beam energy and to agree
with the known perturbative result in that limit.

Our present strategy is to apply a spatial cutoff to the
transverse potential $\chi(\mbox{\boldmath $ \rho$})$ (which has been up
to now set to $2 Z \alpha \ln{\rho}$) in order to obtain an expression
consistent with the perturbation theory formula\cite{bs,kai1} in the
ultrarelativistic limit.  Instead of regularizing the transverse integral
itself Eq. (5) and letting the
cutoff radius go to infinity as was originally done\cite{sw1,bmc,sw2,eg},
we will rather apply an appropriate physical cutoff to the interaction
potential.  In the Weizsacker-Williams or equivalent photon treatment of
electromagnetic interactions the effect of the potential is cut off at
impact parameter
$b \simeq \gamma / \omega$, where $\gamma$ is the relativistic boost
of the ion producing the photon and $\omega$ is the energy of the photon.
If
\begin{equation}
\chi(\mbox{\boldmath $ \rho$}) = \int_{-\infty}^\infty dz V(\sqrt{z^2+\rho^2})
\end{equation}
and $V(r)$ is cut off in such a physically motivated way, then\cite{lm2}
\begin{equation}
V(r)={-Z \alpha \exp[-r \omega_{A,B} / \gamma] \over r }
\end{equation}
where 
\begin{equation}
\omega_A= {p_+ + q_+ \over 2 };\  \omega_B={p_- + q_- \over 2 }
\end{equation}
with $\omega_A$ the energy of the photon from ion $A$ moving in the positive
$z$ direction and $\omega_B$
the energy of the photon from ion $B$ moving in the negative $z$
direction.  Note that we work in a different gauge than that used to obtain
the original perturbation theory formula, and thus our potential picture is
somewhat different.  The transverse potential will be smoothly cut off at a
distance where the the longitudinal potential delta function approximation is
no longer valid.

The integral Eq. (15) can be carried out to obtain
\begin{equation}
\chi(\rho)= - 2 Z \alpha K_0(\rho \omega_{A,B} / \gamma),
\end{equation}
and Eq. (5) is modified to
\begin{equation}
F_{A,B}({\bf k}) = 2 \pi \int d \rho \rho J_0(k \rho)
\{\exp [2i Z_{A,B}\alpha K_0(\rho \omega_{A,B} / \gamma)] -1 \} .
\end{equation}
$F_{A}({\bf k})$ and $F_{B}({\bf k})$ are functions of virtual photon
$\omega_A$ and $\omega_B$ respectively.  The modified Bessel function
$K_0(\rho \omega / \gamma) = - \ln(\rho)$ plus constants for small $\rho$
and  cuts off exponentially at $\rho \sim \gamma / \omega $.  This
is the physical cutoff to the transverse potential.

One may define $\xi = k \rho$ and rewrite Eq.(19) in terms of a normalized
integral $I_{A,B}(\gamma k / \omega)$
\begin{equation}
F_{A,B}({\bf k}) = {4 \pi Z_{A,B} \alpha \over k^2} I_{A,B}(\gamma k / \omega)
\end{equation}
where 
\begin{equation}
I_{A,B}(\gamma k / \omega) =  {1 \over 2 i Z_{A,B} \alpha }
\int d \xi \xi J_0(\xi) \{\exp [2i Z_{A,B}
\alpha K_0(\xi \omega / \gamma k)] - 1 \}.
\end{equation}
It is now clear that $F_{A,B}$ is a function of
$4 \pi Z_{A,B}/k^2$ times a function of $(\gamma k / \omega)$.
The limit as $Z\to0$ of $I^0_{A,B}(\gamma k / \omega)$ is analytically
soluble
\begin{equation}
I^0_{A,B}(\gamma k / \omega) = { 1 \over 1 + \omega^2 / k^2 \gamma^2},
\end{equation}
and
one has $F^0_{A,B}({\bf k})$, the familiar perturbation theory form 
\begin{equation}
F^0_{A,B}({\bf k}) = {4 \pi Z_{A,B} \alpha \over k^2 + \omega^2 / \gamma^2}.
\end{equation}
As I have shown in a previous paper\cite{ajb3}, one might use
some other physical cutoff and still obtain the Lee-Milstein Coulomb
correction as long as one was expanding the Coulomb cross section correction
only to lowest order in $k^2$.  However such an alternate physical cutoff
would not lead to this correct perturbation theory form for
$F^0_{A,B}({\bf k})$ and would lead to modified results for the
Coulomb corrections in a full integration over ${\bf k}$.

Fig. (1) displays the results of numerical calculation
of $\vert I(k \gamma / \omega) \vert^2$ for $Z = 82$ and in
the perturbative limit.  Note that the upper cutoff of $\rho$ at
$\gamma / \omega $
has the effect of regularizing $F({\bf k})$ at small $k$.  $F({\bf k})$ goes
to the constant $4 \pi \gamma^2 / \omega^2$ as $k$ goes to zero in the
perturbative case; it goes to a reduced constant value as $k$ goes to
zero for $Z=82$.  The form of the original solution Eq. (14)
\begin{equation}
F({\bf k}) = {4 \pi \alpha Z \over k^{2 - 2 i \alpha Z} }
\end{equation}
is simply wrong because it is unphysical.  Since it lacks a proper physical
cutoff in $\rho$, it not only blows up at $k = 0$, but it also fails to
exhibit the correct reduction in magnitude that occurs when
$k \gamma / \omega$ is not too large.

%%%%%%%%%%%%%%%%%%%%%%%%%%%%%%%%%%%%%%%%%%%%%%%%%%%%%%%%%%%%%%%%%%%%%%%%%
\begin{figure}[h]
\begin{center}
\epsfig{file=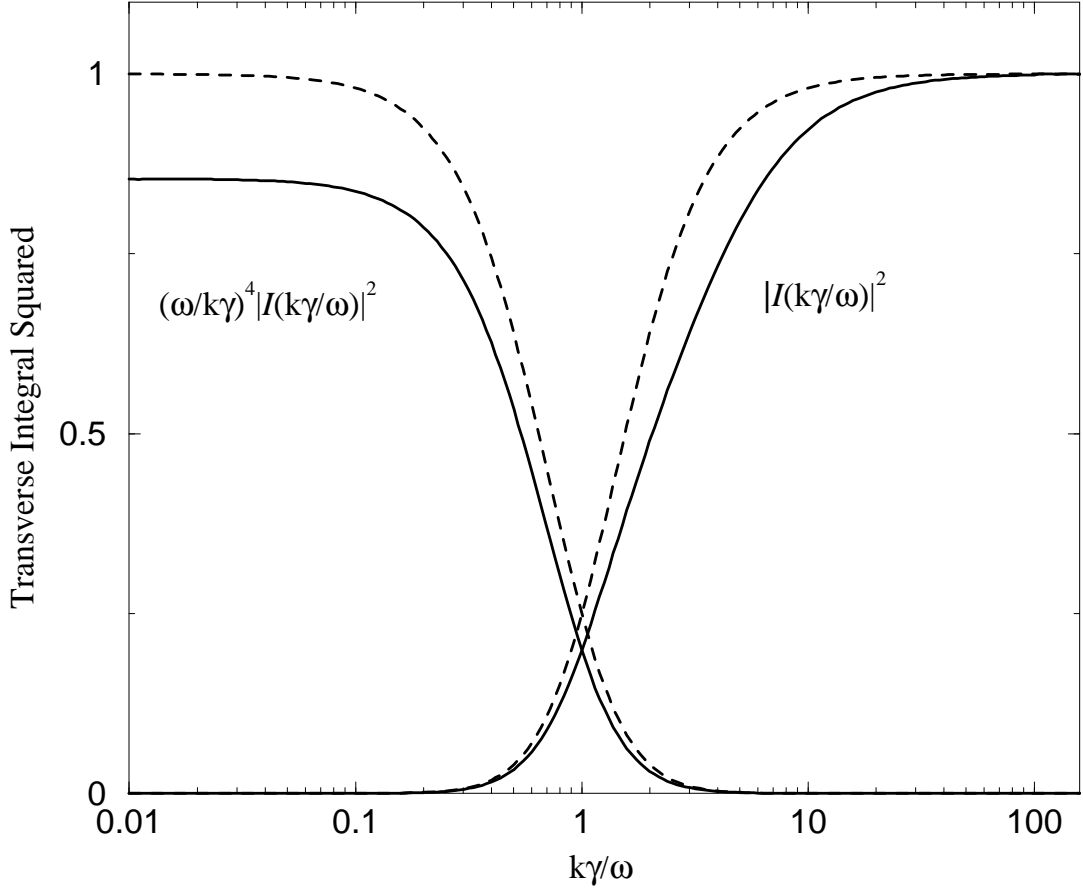,height=12cm}
\end{center}
\caption{The sold curve is the normalized integral squared for Z=82.  The
dashed line is the corresponding perturbation theory result.}
\label{graph1_fig}
\end{figure}
%%%%%%%%%%%%%%%%%%%%%%%%%%%%%%%%%%%%%%%%%%%%%%%%%%%%%%%%%%%%%%%%%%%%%%%

It is clear from Figure 1 that for large
$Z=82$ Coulomb corrections reduce $\vert F({\bf k})\vert^2 $ from the
perturbative result for
$k \gamma / \omega << 100$.  Only for $k >\ \sim 100\ \omega / \gamma $ does 
the magnitude of $F({\bf k})$ go over into the perturbative result.

\section{Calculations:Numerical Technique and Results}

The expression for the total cross section, Eq. (8), involves an eight
dimensional integral over the positron and electron momenta as well as the
virtual photon transverse momentum.  This integral reduces to seven dimensions
in the usual way by symmetry, e.g. let the positron transverse momentum
define the x-axis.  The usual method of evaluation, e.g. in perturbation
theory, is via Monte Carlo.  However, I have chosen to do the seven dimensional
integral directly on meshes uniform on a logarithmic scale in each radial
momentum
dimension.  It was possible to carry the calculation out without using Monte
Carlo because the integrand is very smooth and smoothly goes to zero at both
high end and low end of the momentum ranges.  No artificial cutoffs were
applied.

Calculations labeled exact and perturbative differ only in the
expressions used for $F_{A,B}({\bf k})$.  The analytical expression
Eq. (23) is used for perturbative calculations.  The exact calculations
makes use the expression Eq. (19), which must be evaluated numerically, but
only once for each $Z_{A,B}$ of interest.

Results of the numerical calculations will be compared with previously
derived closed formulas for total cross sections.  It is useful here to
review those formulas.
The Racah formula for the total $e^+ e^-$ cross section in perturbation theory
is\cite{rac}
\begin{equation}
\sigma_R = { (Z_1 \alpha )^2 (Z_2 \alpha )^2 \over \pi m^2}
\biggl[ {28 \over 27} {\cal L^{\rm 3}} - {178 \over 27} {\cal L^{\rm 2}}
%\nonumber \\
+ {370 + 7 \pi^2 \over 27} {\cal L} - {116 \over 9} - {13 \pi^2 \over 54}
+ {7 \over 9} \zeta(3)\biggr]
\end{equation}
where 
\begin{equation}
{\cal L} = \log \biggl[2 {P_1 \cdot P_2 \over M_1 M_2}\biggr]
= [\log 2 (2 \gamma^2 - 1)],
\end{equation}
the relativistic $\gamma$ is that of each colliding ion in an equal and
opposite ion velocity frame, and the Riemann zeta function 
\begin{equation}
 \zeta(3) = \sum_{n=1}^{\infty}{ 1 \over n^3} = 1.2020569 .
\end{equation}
The $\log^3(\gamma^2)$ term is the same as the
original Landau-Lifshitz formula\cite{ll}, but the other additional terms
are an improvement that allows this very early formula to attain a remarkable
degree of accuracy as demonstrated by comparison with recent Monte Carlo
evaluations.

The Lee-Milstein formula\cite{lm2} includes higher order $\alpha Z$ effects
in addition to the $\log^3(\gamma^2)$ term,
\begin{eqnarray}
\sigma_{LM} =&&{(Z_1 \alpha )^2 (Z_2 \alpha )^2 \over  \pi m^2} {28 \over 27}
\biggl[\log^3(\gamma^2) - 3[f(Z_A \alpha)+f(Z_B \alpha)] \log^2(\gamma^2)
\nonumber \\
&&+ 6 f(Z_A \alpha) f(Z_B \alpha) \log(\gamma^2)\biggr]
\end{eqnarray}
where
\begin{equation}
f(Z_{A,B}\alpha) = Z_{A,B}^2 \alpha^2\sum_{n=1}^{\infty}
{ 1 \over n (n^2  + Z_{A,B}^2\alpha^2)}.
\end{equation}
The dominant (negative)
Coulomb correction in this formula is the $\log^2(\gamma^2)$ term,
which was originally obtained by Ivanov, Schiller, and Serbo\cite{serb}
with the Weizsacker-Williams approximation.  
The last (positive) $\log(\gamma^2)$ term in the formula can be thought
of as representing the Coulomb correction corresponding to multiple photon
emission of both ions\cite{lms} and as we show below is relatively small.
%%%%%%%%%%%%%%%%%%%%%%%%%%%%%%%%%%%%%%%%%%%%%%%%%%%%%%%%%%%%%%%%%%%%%%%
\begin{table}
\caption[Table I]{Computer calculations compared with analytical formula
results.  $\gamma$ is defined for one of the ions in the frame of equal
magnitude and opposite direction velocities.  Total cross sections are
expressed in barns.  The positive contribution of multiple photon emission from
both ions to the overall difference betwen exact and perturbative results
is shown in parentheses.}
\begin{tabular}{|cc|ccc|}
\hline
&& Exact & Perturbative & Difference \\
\hline
Pb + Au&Computer Evaluation& 2670 & 3720 & -1050 (+80)  \\
$\gamma = 9.2$&Racah Formula&  &  3470&  \\
&Lee-Milstein& 3050 & 5120 & -2070 (+160) \\
\hline
S + Au&Computer Evaluation& 119.7 & 141.6 & -21.9 (+0.15) \\
$\gamma = 9.2$&Racah Formula&  &  132.0&  \\
&Lee-Milstein& 152.0 & 195.0 & -43.0 (+0.30) \\
\hline
Pb + Pb&Computer Evaluation& 3210 & 4500 & -1290 (+100) \\
$\gamma = 10$&Racah Formula&  &  4210&  \\
&Hencken, Trautmann, Baur&  & 4210 &  \\
&Lee-Milstein& 3690 & 6160 & -2470 (+190) \\
\hline
Au + Au&Computer Evaluation& 28,600 & 34,600 & -6,000 (+220) \\
$\gamma = 100$&Racah Formula&  &  34,200& \\
&Hencken, Trautmann, Baur&  & 34,000 &  \\
&Lee-Milstein& 34,100 & 42,500 & -8,400 (+290) \\
\hline
Pb + Pb&Computer Evaluation& 201,000 & 227,000 & -26,000 (+600) \\
$\gamma = 2960$&Racah Formula&  &  226,000& \\
&Lee-Milstein& 226,000 & 258,000 & -32,000 (+700) \\
\hline
\end{tabular}
\label{tabi}
\end{table}
%%%%%%%%%%%%%%%%%%%%%%%%%%%%%%%%%%%%%%%%%%%%%%%%%%%%%%%%%%%%%%%%%%%%%%%%

Table I shows the results of numerical calculations.  The present
perturbative computer calculations are in good agreement with the Racah
formula at RHIC and LHC energies as expected, and
with the published Monte Carlo RHIC calculations of Hencken, Trautmann, and
Baur\cite{kai,AlscherHT97}.  At SPS energies the present perturbative computer
calculation results are a bit higher (7\%) than the Racah formula and the
Hencken, Trautmann, and Baur calculation, perhaps indicating divergence
in those results from the ultrareletivistic limit of the present treatment.
The full numerical evaluation of the exact semi-classical
total cross section for $e^+ e^-$ production with gold or lead ions
shows reductions from perturbation theory of 28\% (SPS), 17\% (RHIC),
and 11\%(LHC).  Clearly with increasing beam energy (and a larger value
for the spatial cutoff of the transverse integral in the formula)
higher order corrections to perturbation theory are relatively smaller.
The S + Au calculation at SPS energy shows an expected smaller reduction
from perturbation theory (15\%) than the 28\% reduction of Pb + Au at
the same energy.
The Lee-Milstein higher order overall correction to perturbation theory
(difference column) is
negative but somewhat larger than the difference evaluated here numerically.  
The small positive contribution of multiple photon emission from both ions
to the overall negative Coulomb correction is shown in parentheses
in the difference column.  Because of the way the numerical calculations
were organized it was straightforward to extract this contribution
from the exact computer evaluation.  Again the Lee-Milstein formula
overestimates this small positive contribution, especially for the
SPS case.

%%%%%%%%%%%%%%%%%%%%%%%%%%%%%%%%%%%%%%%%%%%%%%%%%%%%%%%%%%%%%%%%%%%%%%%%%
\begin{figure}[h]
\begin{center}
\epsfig{file=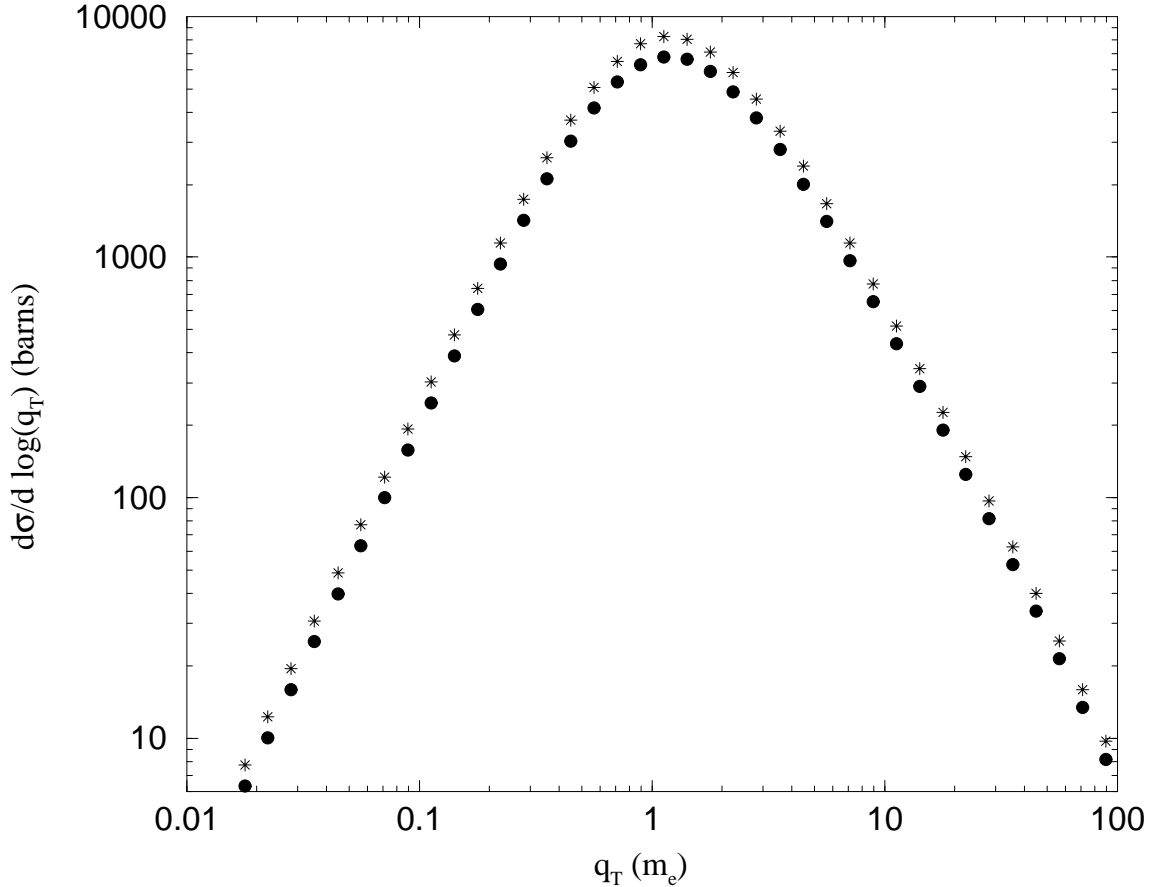,height=12cm}
\end{center}
\caption{Positron transverse momentum spectrum for Au + Au at RHIC
with $\gamma = 100$.  The filled circles are the exact calculation
and the stars the perturbation theory.}
\label{graph2_fig}
\end{figure}
%%%%%%%%%%%%%%%%%%%%%%%%%%%%%%%%%%%%%%%%%%%%%%%%%%%%%%%%%%%%%%%%%%%%%%%

There is the question of whether Coulomb corrections might become
vanishingly small in some momentum regions.  Let us take the Au + Au
RHIC case as an example.  If one looks at the uncorrelated positron
cross section cross section (Eq. (7)) as a function of momentum then
one finds that throughout the transverse and longitudinal momentum space
of the final positron, the smallest reduction from perturbation theory
is 12.5\% and the largest reduction is 25\% in comparison to the
mean or integrated total cross section reduction of 17\% of the table.
Thus the argument given in Ref. \cite{serb} that Coulomb corrections
contribute mostly for ${\bf q_{\perp} = m_e}$, but should disappear for
larger and smaller ${\bf q_{\perp}}$ is not verified.

Figure 2 shows the transverse momentum spectrum integrated over all
longitudidal momenta.  The overall contribution does peak at about
${\bf q_{\perp} = m_e}$.  However,
Coulomb corrections persist to the highest and lowest values of
${\bf q_{\perp}}$, scaling roughly with the perturbative cross section.
Figure 3 shows the logitudinal momentum spectrum
integrated over all transverse momenta, and likewise Coulomb corrections
persist to the highest and lowest values of $q_z$.

%%%%%%%%%%%%%%%%%%%%%%%%%%%%%%%%%%%%%%%%%%%%%%%%%%%%%%%%%%%%%%%%%%%%%%%%%
\begin{figure}[h]
\begin{center}
\epsfig{file=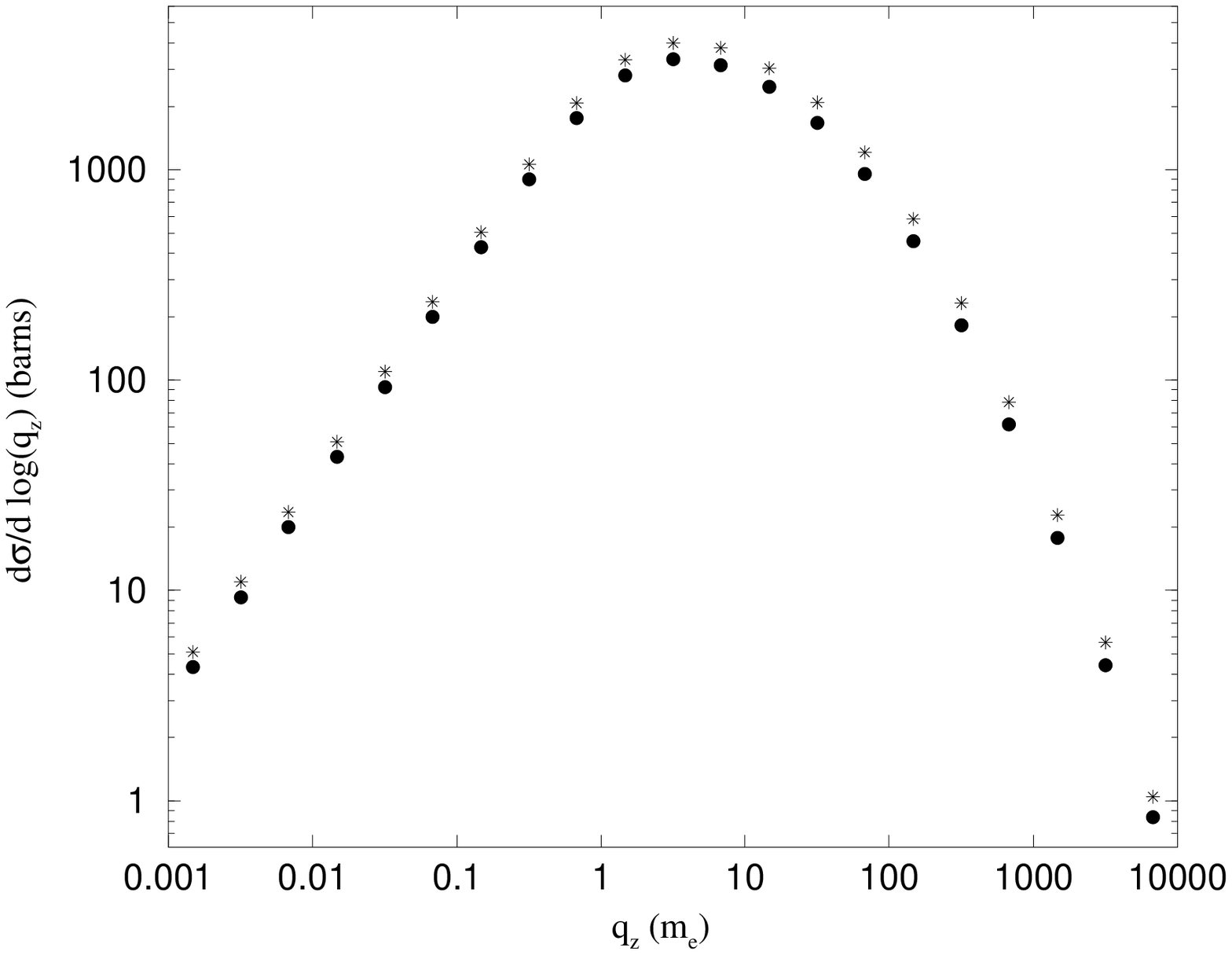,height=12cm}
\end{center}
\caption{As in Figure 2 but for the positron longitudinal momentum spectrum
for Au + Au at RHIC with $\gamma = 100$.}
\label{graph3_fig}
\end{figure}
%%%%%%%%%%%%%%%%%%%%%%%%%%%%%%%%%%%%%%%%%%%%%%%%%%%%%%%%%%%%%%%%%%%%%%%

Given the decrease of Coulomb corrections with increasing beam energy
one might ask, ``At what $\gamma$ of colliding Pb beams would Coulomb
corrections be relatively unimportant, say, less than 1\% for the total
cross section?'' If for the purposes of {\it reductio ad absurdum}
one takes the Lee Milstein formula as a reasonable order of magnitude
approximation, then the answer is $\gamma=10^{43}$.  The point is that
for any conceivable
accelerator beyond LHC the Coulomb corrections to $e^+ e^-$ pair production
will still be significant.

One can calculate momentum spectra to compare with the CERN SPS data.
Since the CERN data comprise
positrons uncorrelated with electrons, comparison with a full calculation
of the positron momentum spectrum $d\sigma(q)$ is appropriate.  Figure 4
shows the data for a Pb projectile
on a Au target.  On the whole the perturbation theory curve (dashed line)
perhaps seems closer to the data than to the solid full exact
calculation.

%%%%%%%%%%%%%%%%%%%%%%%%%%%%%%%%%%%%%%%%%%%%%%%%%%%%%%%%%%%%%%%%%%%%%%%%%
\begin{figure}[h]
\begin{center}
\epsfig{file=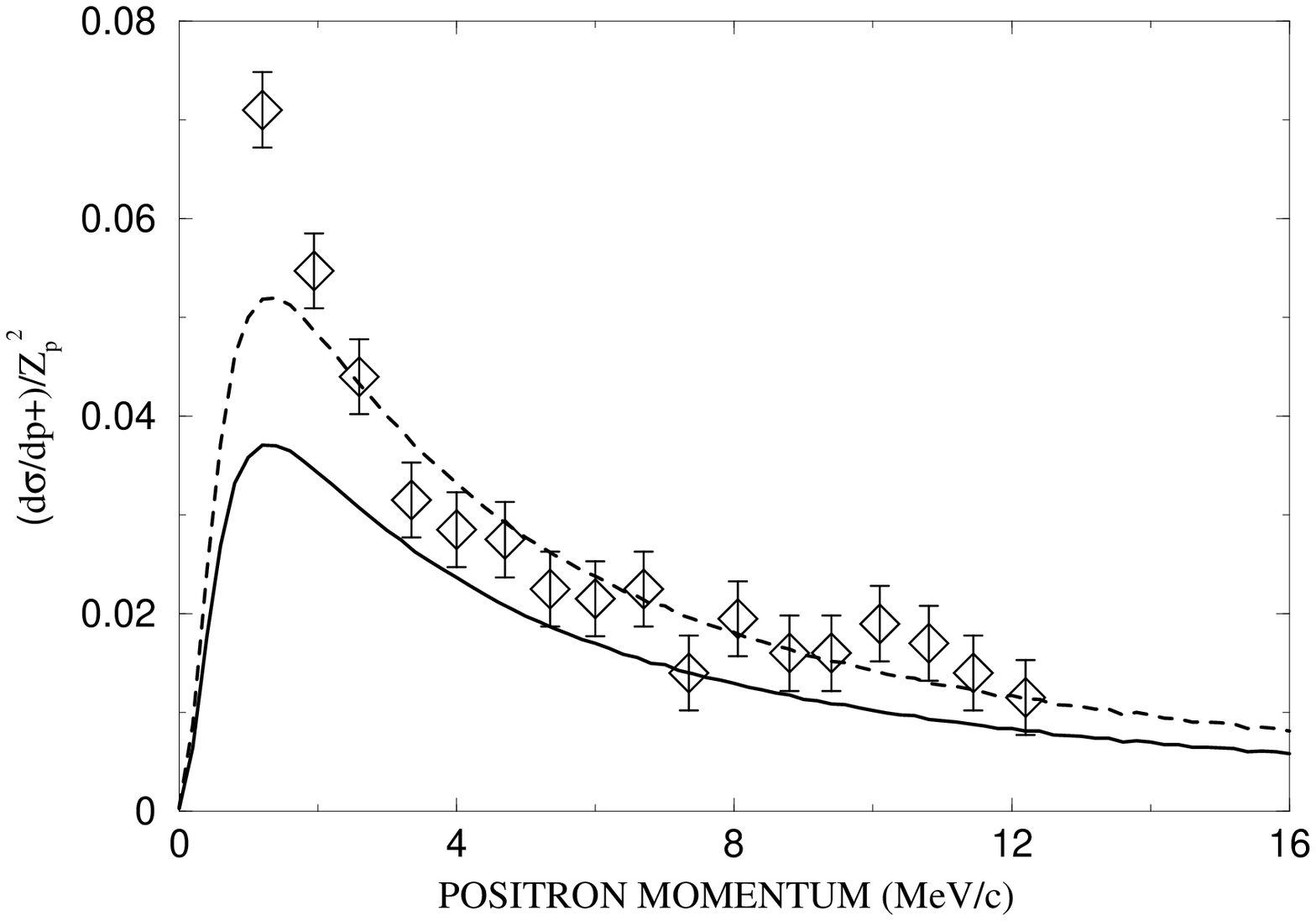,height=12cm}
\end{center}
\caption{Calculated positron momentum spectrum
compared with the CERN SPS data for Pb + Au.  The solid line is the
exact calculation and the dashed line perturbation theory.}
\label{graph4_fig}
\end{figure}
%%%%%%%%%%%%%%%%%%%%%%%%%%%%%%%%%%%%%%%%%%%%%%%%%%%%%%%%%%%%%%%%%%%%%%%

Figure 5 show an analagous comparison for a S projectile on a Au target.
Again, the perturbation theory curve seems closer to the data, represented
by the dot-dashed line.  Figures 4 and 5 provide an illustration of the
statement of the experimental authors, that the cross sections follow
perturbative scaling.  However, especially given the difficulty of the SPS
experiment as described by the authors, the apparent lack of Coulomb
corrections seen here needs to be verified in other ultrarelativistic
heavy ion experiments.

%%%%%%%%%%%%%%%%%%%%%%%%%%%%%%%%%%%%%%%%%%%%%%%%%%%%%%%%%%%%%%%%%%%%%%%%%
\begin{figure}[h]
\begin{center}
\epsfig{file=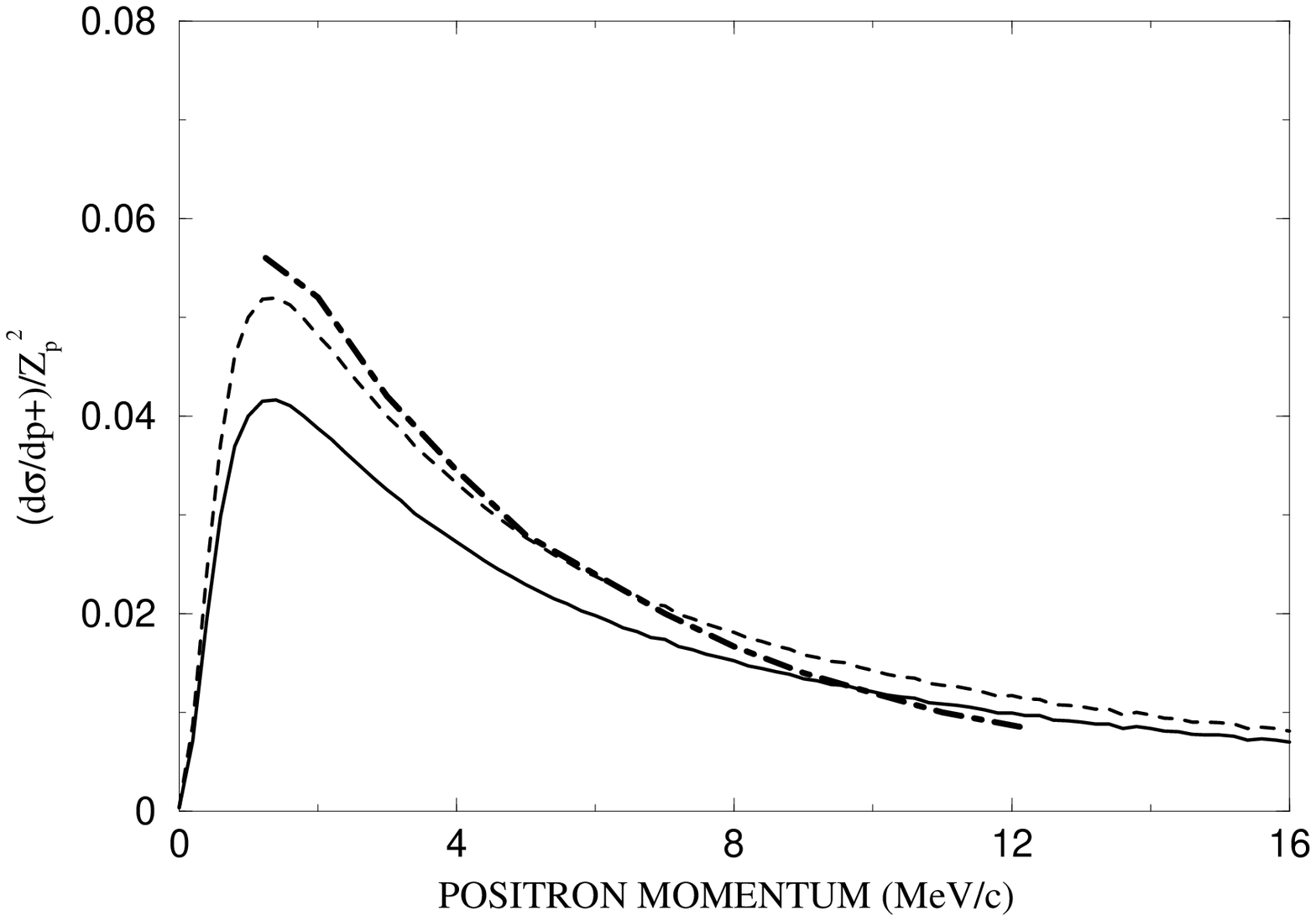,height=12cm}
\end{center}
\caption{As in Figure 4 but with a S projectile.  The dot-dashed line
follows the authors' representation of the CERN SPS data.}
\label{graph5_fig}
\end{figure}
%%%%%%%%%%%%%%%%%%%%%%%%%%%%%%%%%%%%%%%%%%%%%%%%%%%%%%%%%%%%%%%%%%%%%%%

The first experimental observation of $e^+ e^-$ pairs at RHIC has been
published by STAR\cite{star04}.  Events were recorded where pairs were
accompanied by nuclear
dissociation.  Comparison with perturbative QED calculations allowed
a limit to be set ``on higher-order corrections to the cross section,
$-0.5 \sigma_{QED} < \Delta \sigma < 0.2 \sigma_{QED}$ at a 90\% confidence
level.''

The present technology of the properly regualarized exact computer
code does not include an impact parameter representation and thus does not
allow for evaluating a cross section where pair production is in
coincidence with nuclear dissociation.  Furthermore, the retarded propagators
are not strictly appropriate when the range of both electrons and positrons
are restricted such as the STAR data.  However, a comparison of calculations
in the STAR acceptance without nuclear dissociation is of interest as an
indication
of the relative difference between perturbation theory and the regularized
exact result.  In the STAR acceptance the exact result is calculated to
be 17\% lower than perturbation theory (as is coincidently true for the
total RHIC e+e- cross section in Table I).  This rough estimate
$ \Delta \sigma = -0.17 \sigma_{QED} $ is not excluded by the above STAR limit.
 
\section{Summary and Conclusions}

A full numerical evaluation of the ``exact'' semi-classical
total cross section for $e^+ e^-$ production with gold or lead ions
shows reductions from perturbation theory of 28\% (SPS), 17\% (RHIC),
and 11\%(LHC).

For large Z no final momentum region was found in which there was
no reduction or an insignificant reduction of the exact cross section
from the perturbative cross section.

The CERN SPS data cover a large part of the momentum range of produced
positrons, and the present theory predicts a reduction of cross section
at high Z from the perturbative result.  That the CERN SPS data
apparently do not show a reduction from perturbation theory is a puzzle.
It would be of great interest to obtain more precise data on variation of
heavy ion pair production cross sections with ion charge at RHIC or LHC.
If the present apparent lack of evidence for Coulomb corrections
in ultrarelativistic heavy ion $e^+ e^-$ pair production were to be
reproduced in other experiments it would provide a unique challenge to
our theoretical understanding of strong field QED. 

\section{Acknowledgments}
I would like to thank Francois Gelis, Kai Hencken, Larry McLerran,
Valery Serbo, and
Werner Vogelsang for useful discussions.  This manuscript has been authored
under Contract No. DE-AC02-98CH10886 with the U. S. Department of Energy. 
%\vfill\eject

\end{document}